\documentclass[conference]{IEEEtran}
\usepackage{cite}
\usepackage{amsmath,amssymb,amsfonts}
\usepackage{algorithmic}
\usepackage{graphicx}
\usepackage{textcomp}
\usepackage{xcolor}
\usepackage{tabularx}
\def\BibTeX{{\rm B\kern-.05em{\sc i\kern-.025em b}\kern-.08em
    T\kern-.1667em\lower.7ex\hbox{E}\kern-.125emX}}
\begin{document}

\title{Ambiguity in Utopian XR-Games. Basic Principles for the Design of Virtual Worlds\\
}

\author{\IEEEauthorblockN{1\textsuperscript{st} Wolfgang H{\"o}hl}
\IEEEauthorblockA{\textit{Department of Informatics} \\
\textit{Technical University of Munich (TUM)}\\
Munich, Germany \\
wolfgang.hoehl@tum.de}
}
\maketitle

\begin{abstract}
Utopian images in XR-games are often ambiguous. How can ambiguity be consciously designed in virtual worlds? What are the design principles for game designers? Ambiguity arises from discontinuity and decontextualization, from the deliberate omission of facts, and from the skillful superimposition of contradictory codes. An ontology of interactive media explains the elusive visual poetry and image perception. The five basic elements of media are presented. The model shows the key correlations of user, content, and technology in image perception. An overview of utopian representation as a cross-media phenomenon places game design in an interdisciplinary context. Six selected computer-generated illustrations show how the above design criteria can be successfully applied to the design of virtual worlds. The ambiguous utopia of an image exists only in a brief, irretrievable moment of our own perception. But this moment is valuable for designing real visions for a better future.\\
\end{abstract}

\begin{IEEEkeywords}
utopian image, ambiguity, image perception, image design, game design, game development.
\end{IEEEkeywords}


\section{The Current Present is the Real Utopia}

"Reality is boring" said one student when I asked him why he particularly likes XR-games. Of course, these computer games are very attractive. They catapult us into three-dimensional virtual worlds. They mirror a dream world for us, where we can take on fantastic roles, defeat overpowered opponents and discover places we've never seen before. We can feel "sublime" or "epic" for a brief moment. Sublimity and epic feelings - today often tortured words. Why, this is perhaps something to think about more deeply. XR-games lend themselves to escapism and procrastination, an escape from a reality that is perhaps too complex, incomprehensible and perceived as too boring.

In fact, however, this escape does not succeed. It never succeeds. Escape is impossible. Games are part of our reality. According to Popper this reality comprises three worlds \cite{b1}: the physical, the psychological and the spiritual world. When we play, we are always dealing with reality. In XR-games, it is predominantly our psychological and spiritual reality. Multi-layered ambiguities fire our psyche and electrify our mind. Fictional roles, captivating stories and fantastic universes are the illusions that seemingly free us from the annoying physical world. Often we forget about the fact that we are actually only concerned with ourselves. This is very comfortable. Ambiguities throw us back on ourselves much more. However, the basic technology and we ourselves always remain part of the relentless physical world. This only becomes clear to us when something doesn't work. Syntax Error. I need a coffee.

Utopia is also always concerned with reality. That may sound paradoxical, but it's true. Utopia cannot be detached from our tripartite reality. Why then are utopian ideas so appealing to us? They appeal to our inner selves. They activate our own imagination. And they remain multilayered and ambiguous. Thus, they remain open to our dreams, desires, and fantasies. As well-packaged reality, they criticize current conditions and open up a space for us to dream. A fantasy world that lies in the distant past. Or a future that takes place on an unknown planet. Utopia is always far away - in time or space. It is unattainable, yet at the same time always present in the present moment, in our imaginations, thoughts, fantasies and dreams. Utopia lives from this intrinsic paradox. Forever.

This paper explores the multi-layered ambiguity of utopian environments in XR-games. Paradoxically, ambiguity and utopia lead us back to real life in interesting detours. Ambiguity activates our psyche and mind. The timeless and placeless utopia manifests itself exclusively in a brief moment of our present consciousness. The present and reality are the real utopia. XR-games and virtual worlds are a real chance to depict our present conditions and to understand deficiencies. Thus they are also a means to design our everyday reality sustainable and worth living in all areas.

To treat the whole idea of utopia would go beyond the scope of this paper. It is a long story. Therefore, this work is limited to the utopian representation as needed to develop XR-games. The thesis provides a cross-media overview of utopian representations.

A brief ontology of media explains image perception and the elusive visual poetry. Popper's 3-world model \cite{b1} and the five basic elements of interactive media are introduced \cite{b2}. An extended and improved model for image perception shows important correlations between user, content and technology \cite{b3}. Specific examples are used to elaborate three key principles for ambiguous image perception - discontinuity and decontextualization, omission of facts, and superimposition of contradictory codes. Six computer-generated illustrations show how these principles can be implemented creatively in a concrete image composition.

This work is not only aimed at game designers and game developers. The design principles and theoretical models presented here are meant to be a guideline. They should make it easier for a broad interdisciplinary readership to recognize, evaluate and further develop essential design elements of games in the socio-cultural context of the media. This contribution should help to better classify terms when working with games and to better understand important correlations. On this basis, it should become easier for all designers to better apply the basic principles and to further develop ambiguous representations for XR-games.

\section{The Utopian Image}

\begin{figure*}[htbp]
\begin{minipage}[b]{1.0\textwidth}
\centerline{\includegraphics[width=1.0\textwidth]{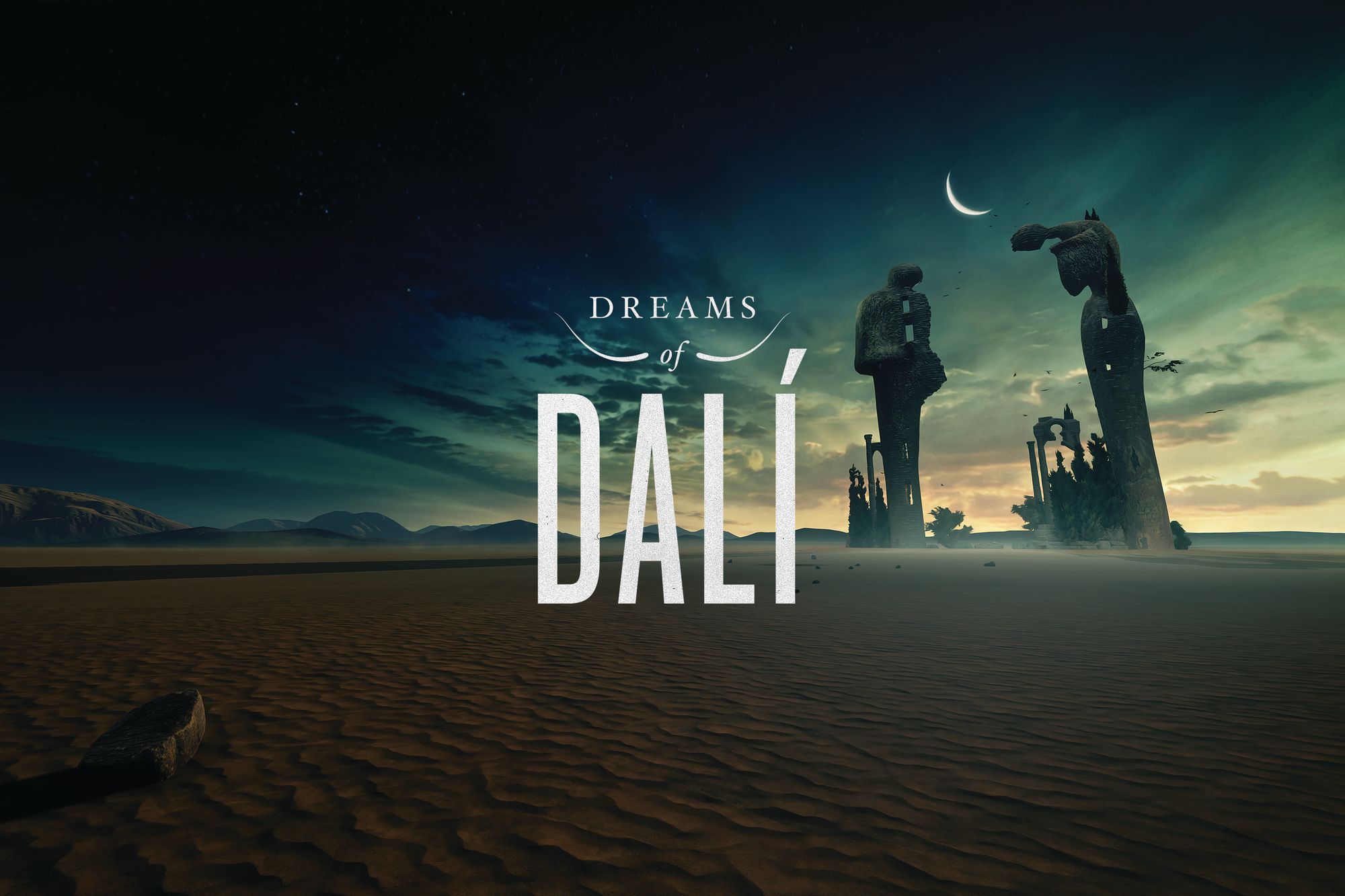}}
\caption{Still image from Dreams of Dalí Virtual Reality (VR) Experience (Salvador Dali Museum, 2016).}
\label{fig1}
\end{minipage}
\end{figure*}

Looking for current XR-games in the utopian genre, there is a lot of commercial but unfortunately low demanding from mainstream entertainment  \cite{b4,b5,b6}. Among them are also refreshingly amateurish indie games \cite{b7} and big titles, such as Mass Effect Andromeda, Dragon Age Inquisition \cite{b8,b9} Skyrim VR \cite{b10} or Grand Theft Auto, The Witcher \cite{b11} and Resident Evil \cite{b12}. In terms of design, I was personally most impressed by Dreams of Dalí in Virtual Reality \cite{b13} as shown in Figure \ref{fig1}. The following section is intended to help place current game developments in a larger, interdisciplinary context, to classify them, and to understand them better. The following overview makes no claim to completeness.

Utopian representation goes back a long way. It is a cross-media phenomenon and can be found in all types of media - theater and performance, painting and print media, film, photography \cite{b14}, electronic and digital media \cite{b15}. Comics, graphic novels, and magazines also feature utopian representations. Boing Boom Tschak \cite{b16}. And last but not least, they are a central issue in architecture and urban planning \cite{b3, b17}.

They are social, religious and technical utopias, showing strange mythical creatures and landscapes, promised and lost paradises. They appear in drawings, paintings, and etchings, such as those by Hieronymus Bosch, Jacques Callot, Lucas Cranach, Johann Heinrich Füssli, and Alfred Kubin. Raffael \cite{b18} developed the transfiguration of Christ. Many motifs of the Surrealists and Dadaists, such as Max Ernst, Salvador Dalí, René Magritte or Giorgio de Chirico are treating the dream interpretation of psychoanalysis, the unconscious and the absurd. The theme also preoccupies the Fantastic Realists around Arik Brauer, Ernst Fuchs, Rudolf Hausner, Friedensreich Hundertwasser or Anton Lehmden. There are also illustrations in books by Thomas Morus \cite{b19, b20}, Heinrich von Kleist \cite{b21}, Jules Verne \cite{b22, b23, b24} Herbert George Wells \cite{b25}, Burrhus Frederic Skinner \cite{b26}, Aldous Huxley \cite{b27}, George Orwell \cite{b28, b29}, Stanislaw Lem \cite{b30, b31, b32}, Ray Bradbury \cite{b33}, Isaac Asimov \cite{b34}, William Gibson \cite{b35} or Douglas Adams \cite{b36}. The list could be continued with Herbert W. Franke, J. G. Ballard, Algernon Blackwood, Marcus Hammerschmidt, Arkadi and Boris Strugatzki, and last but not least H. P. Lovecraft. In Amazing Stories, Hugo Gernsback \cite{b37} graphically revived the works of Jules Verne and Edgar Allen Poe. Perry Rhodan and Flash Gordon followed, right up to Superman, Batman and Marvel's heroes of today. And who doesn't know "Metropolis" by Fritz Lang (1927), "2001 - A Space Odyssey" by Stanley Kubrick (1968) \cite{b38}, "Soylent Green" by Richard Fleischer (1972), "Brazil" by Terry Gilliam (1985), the "Terminator" series or "Avatar" by James Cameron (1984 and 2009), "War of the Worlds", "Star Wars" and "Star Trek", to name but a few. "Retro-Futurism" \cite{b39} shows the "World of Tomorrow" in prints and illustrations by Klaus Bürgle, Erik Theodor Lässig, Kurt Röschl, Eberhard Binder-Staßfurt, Hans and Botho von Römer, Günter Radtke, Helmuth Ellgaard, Heinz Hähnel and Oswald Voh.

The utopian designs of Claude-Nicolas Ledoux and Étienne-Louis Boullée accompany the social upheaval of the French Revolution. The classicist architecture of Canova, Thorwaldsen or Baltrard, on the other hand, seeks utopia in classical antiquity. Haussmann shaped the new cityscape of Paris in the 19th century. The so-called "utopians," such as Robert Owen, Charles Fourier and Etienne Cabet, design new ideal cities for industrialized society. The modernist movement experiments with new building materials, standardized production methods and new forms of housing. The well-known housing estates of the interwar period are created, not only in Stuttgart, Vienna and Berlin. The Italian Futurists around Enrico Prampolini and Antonio Sant'Elia draw megalomaniac traffic junctions and power plants \cite{b40}. In the young Soviet Union, designs for skyscrapers, housing communes, kindergartens, clubs, sports stadiums, linear cities, bathhouses, trade fair pavilions, power plants, and houses of the future are created by architects such as El Lissitzky and Kasimir Malevich \cite{b41}. In the second half of the 20th century, William Katavolos designs chemical-organic building materials for self-growing and adaptable houses and cities \cite{b42}. Konrad Wachsmann, Yona Friedman, Paolo Soleri, Walter Jonas, and Noriaki Kurokawa draw and realize urban megastructures \cite{b43}. Buckminster Fuller designs and manufactures his self-supporting domes, the "Dymaxion" houses, and develops a "world planning program" \cite{b44}. Parallel to the first space flights and the moon landing, Archigram's "Walking Cities" and "Living Pod" are created, along with many other technology- and future-inspired futuristic ideas \cite{b45}. Hans Hollein \cite{b46} presents his "aircraft carrier in the landscape." Even floating cities and settlements on the ocean floor suddenly seem possible \cite{b47}.

What are the utopias of today? According to Jens Jessen \cite{b48}, utopias of today are not directed toward a better social order, nor toward a more equitable distribution of resources, nor toward the liberation of oppressed peoples or classes. Rather, utopias of today are directed toward liberation from human nature. As examples, Jessen cites birth control, "social freezing," the self-optimization and hybridization of humans through genetics, bio-, nano-, and computer technology. He asks about the new role models in the gender debate and also mentions the discussion about life-prolonging technologies, euthanasia and control over the time of one's own death.

It is always current events that give rise to utopias. Utopia sharpens our view of current reality. It processes current developments and shows possibilities how we can better shape our future world. Utopia is not an escape into our own psycho-spiritual world. It is an appeal to set out into a future that we can take into our own hands and shape.

\section{Visual Poetry and Image Perception}

"Remembering is ... an imaginative reconstruction, ..." \cite{b49}.\\

This section presents a draft ontology of media. This ontology is intended to help game designers and other designers better classify concepts in game design and recognize important correlations. Karl R. Popper \cite{b1} often uses the model of three worlds. According to Popper, we can assign all observable phenomena to three different worlds. World 1 is the material world of physical objects and states. World 2 is the world of our perception and consciousness. World 3 is the world of mental products and contents. These can be theories, models, geometric or mathematical theorems. The model is very descriptive, exact and powerful. Therefore, it is also used here to explain the perception of images.\\

"Visual poetry is a kind of catch-all where ambiguity is enhanced by a sort of lyricism that defies explanation." \cite{b50}.\\

Stuart Franklin describes that the ambiguity of an image can be enhanced by reflection and by an elusive visual poetry. Now, how can these ideas be more accurately described in a model?

\begin{figure*}[htbp]
\begin{minipage}[b]{1.0\textwidth}
\centerline{\includegraphics[width=1.0\textwidth]{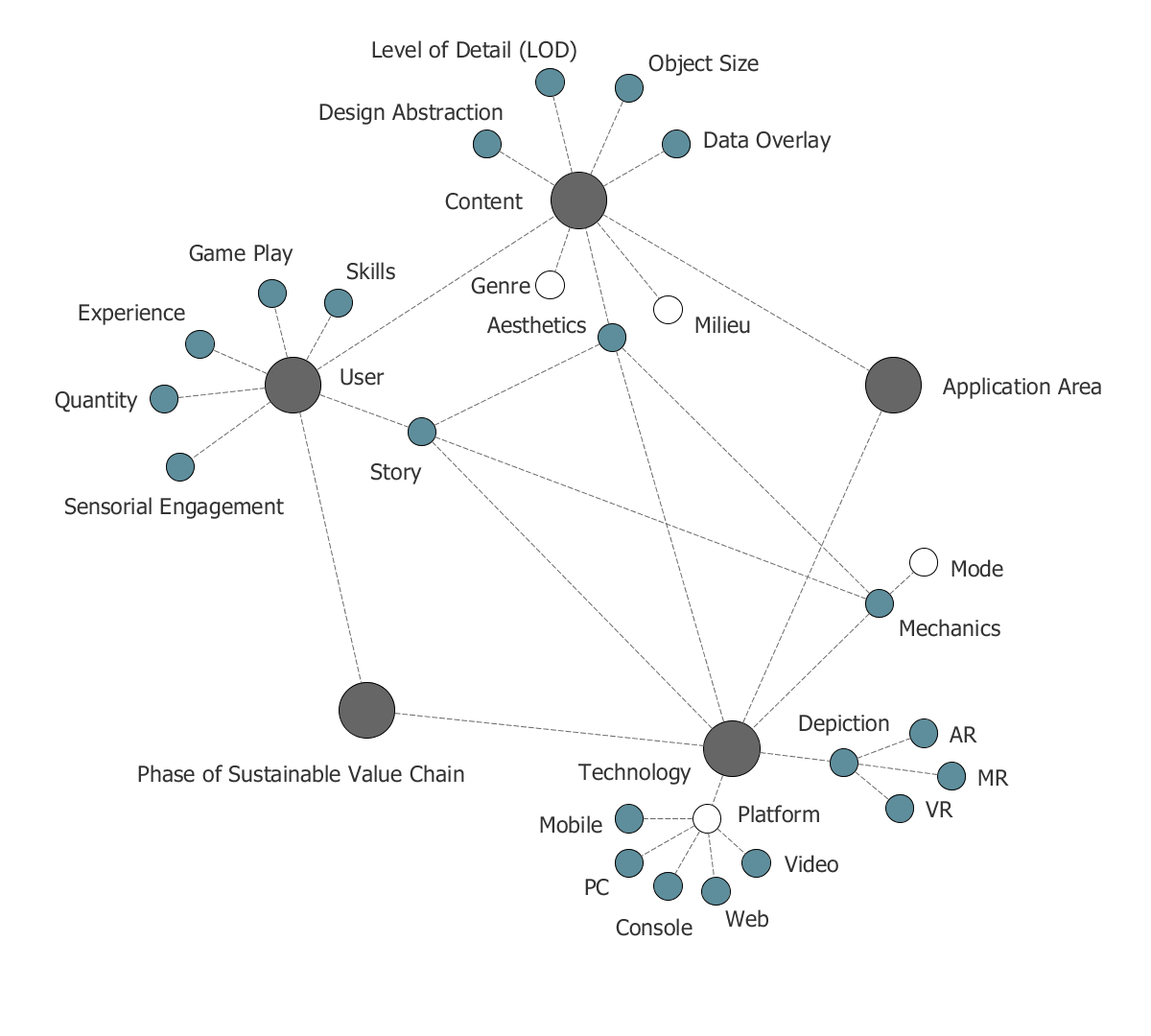}}
\caption{Five basic elements of interactive media (Höhl, 2019).}
\label{fig2}
\end{minipage}
\end{figure*}

Assume that all types of media consist of five basic elements, as shown in Figure \ref{fig2}: user, application area, content, phase of value chain and technology \cite{b2}. This assumption is conform with the four known main elements of a game according to Apperley \cite{b51}: platform, genre, fashion and milieu. The four main elements according to Apperley are drawn as white nodes in Figure \ref{fig2}. These five basic elements also include the four basic components of a game according to Jesse Schell \cite{b52}: aesthetics, story, game mechanics and technology. In Figure \ref{fig2} you can easily see the rhombic arrangement of these four elements. Schell has ordered his four basic components according to visibility. At the top and immediately visible to the player is aesthetics, then come story and game mechanics. Technology remains invisible.

The individual elements can be assigned to Popper's three worlds. The content, aesthetics, story, genre, and milieu are part of world 3. User, technology, value chain, and scope are part of world 1. The user focuses on world 2 - his world of perception and awareness (game play, user experience, sensorial engagement). The user's world 2 brings together world 3 and world 1. Interactive media such as XR games are thus technical objects that fully integrate the user - mentally, psychologically and also physically. This opens up an exciting field between environment, medium and user. The viewer becomes an indispensable and integral part of the interactive media system.

\begin{figure*}[htbp]
\begin{minipage}[b]{1.0\textwidth}
\centerline{\includegraphics[width=1.0\textwidth]{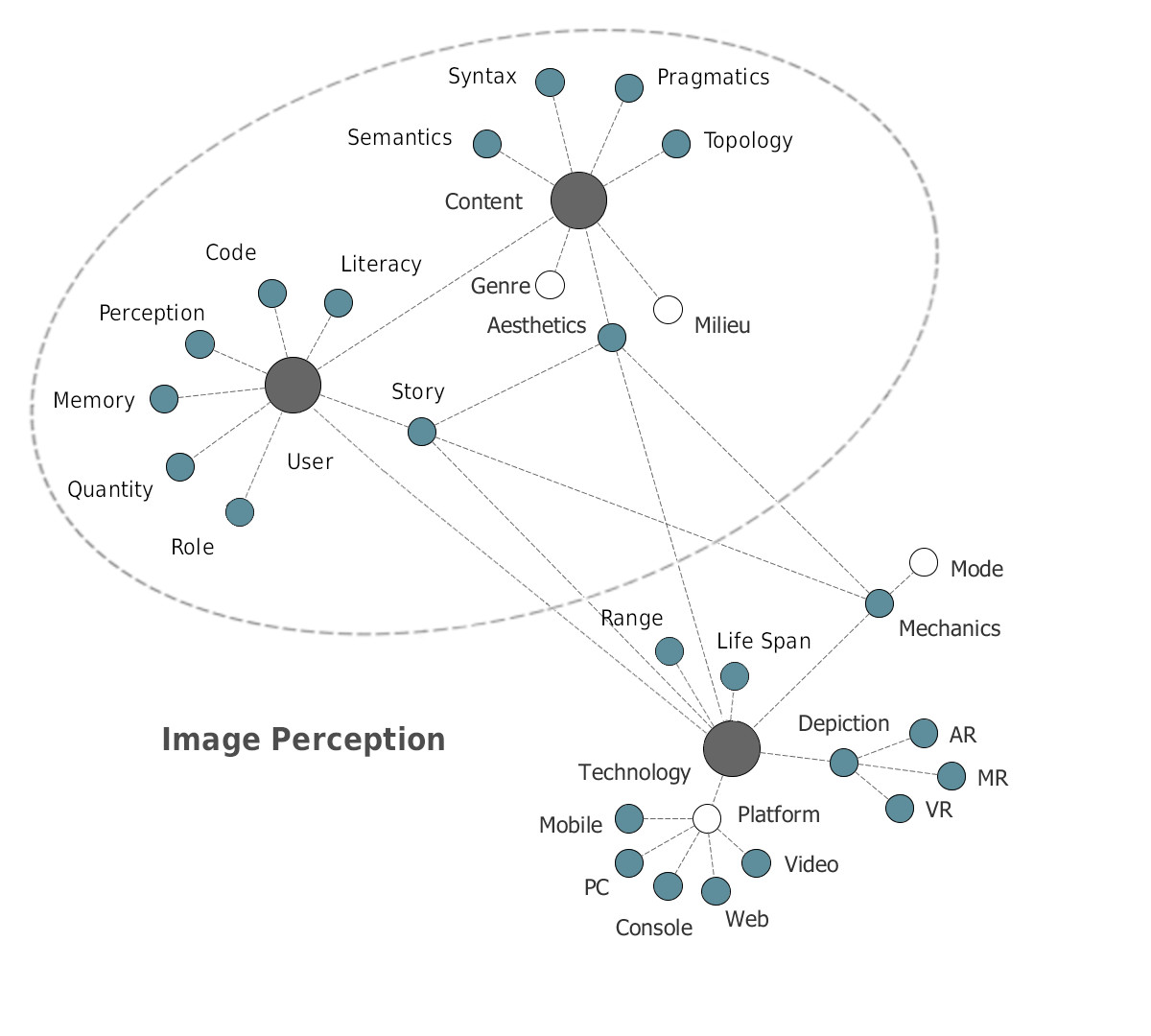}}
\caption{Image perception. Correlations of user, content and technology.}
\label{fig3}
\end{minipage}
\end{figure*}

Figure \ref{fig3} shows a diagram for image perception with the correlations between user, content and technology. Value chain and application domain have been removed from the diagram due to lack of relevance and for more clarity. The technology has a specific depiction method. The life span of the technology and the range have been added. Depiction method and life span determine the range of a medium and thus the possible number of users over time. In linguistics, a distinction is made between syntax, semantics and pragmatics \cite{b53}. With the content of media, the character set (syntax) and the topology were added. We encounter the element of topology again later, in the emergence of ambiguity and ambivalence. The content can be assigned a specific meaning (semantics) for a specific user group (pragmatics). The qualifications or skills of the user are decisive in this context. These include knowledge of the syntax code (literacy) and the ability to perceive and remember (perception, memory). The number of users can vary (quantity). The user also has different roles. They can be authors, producers or recipients of media. In the case of interactive media, there is also the concept of the "produser", a mixture between producer and recipient. Media can be received by a different number (quantity) of users.

The visual poetry of a representation arises gradually differently through the perception, memory and reflection of the viewer. Depending on the individual constitution and on knowledge and experience, the image is perceived. Individual associations arise. This overall perception can be unambiguous, ambiguous or ambivalent.

\section{How does Ambiguity arise?}

How does ambiguity arise in this tense interplay? Franklin \cite{b50} names three criteria for ambiguity in photography:\\

\begin{itemize}
\item Discontinuity and decontextualization.
\item Deliberate omission of facts
\item Superimposition of contradictory codes\\
\end{itemize}

According to Franklin, ambiguity arises from a creative combination of these three criteria. It is created by decontextualization, by embedding the content of the image in a different context. Decontextualization appears in multiple ways in utopian representation. Decontextualization concerns the macro context of the media image. The representation is always in a particular spatial or temporal context. Decontextualization occurs by dislodging or questioning this given context, or by accidental recontextualization. This can be, for example, an incomprehensible syntax and semantics of the representation, an insufficient pragmatics or an inappropriate aesthetics. Or the image may be associated with a seemingly inappropriate or random text. It is the play of chance.

Ambiguity also arises from discontinuity, a break in the image narrative. This involves a micro-context within the image itself. Discontinuity happens in the content of the picture. Image objects that seemingly do not fit together, a surprising topology. This also includes the deliberate omission of facts. The viewer is left in the dark about the history and background of the picture. Topology is a central starting point for ambiguous representations.

Hollein \cite{b46} uses discontinuity and decontextualization in his "Aircraft Carrier in the Landscape." He combines two pictorial objects that are unambiguous in themselves in a new, confusing context. Preuss \cite{b54} also leaves the viewer uncertain where his design of a visionary city for Mass Effect Andromeda is actually located. He combines a utopian cityscape with a gigantic climate shell. It's an oversized, sunlit glass wall and a huge ventilation system that lights and ventilates the city. Discontinuity and decontextualization is also the theme of the monolith in Kubrick's "2001 - A Space Odyssey" \cite{b38}. It is suddenly there. Without any prehistory, it suddenly appears in the context of a possible future. It is like the awakening of Gregor Samsa as an insect in Franz Kafka. Something disturbs the usual topology of the narrative.

In the context of the perception of spatial topology, Deleuze and Guattari \cite{b55} develop the two concepts of "smooth" and "notched" space. In their view, "smooth" space is "infinite, open, and unbounded in all directions" and "spreads a continuous variation." The authors describe an open space for the unpredictable and free action of users. The "notched" space is ordered and structured for them. There are no surprises there, all events are potentially determined and predictable. They see both concepts of space as unattainable ideals that do not exist in real terms on their own. They describe "smooth" and "notched" space as polar opposites. They describe open space as "a zone of unpredictability inherent in becoming." Open space lies "somewhere in between" - between the ideals of "smooth" and "notched" space. For the authors, both types of space are always intimately intertwined and "drive each other forward"  \cite{b55}. It is the present simultaneity of different intensities of the two opposing concepts that enables ambiguous and meaning-open spaces. According to Tulatz \cite{b56} notched spaces strive for contingency reduction and smooth spaces for contingency expansion1. Perceived spatial topology enables or denies contingent events.

Tschumi and Ruby \cite{b57} introduce the notion of spatial topology. The topology of a space can be perceived isotropic or anisotropic. Or somewhere in between. The open space, the open representation needs the simultaneous presence of different intensities of polar concepts: simultaneous isotropy with anisotropy present, synchronously present smooth and notched space, contingency expansion with simultaneous contingency reduction. Above all, it needs a certain margin of indeterminacy, an open concept, to allow the user's appropriation of space, to include all users, to fully integrate them into the open space.

Ambiguity means that both pairs of opposites are present at the same time when one thinks of even one of the two terms. When thinking of a distant land or a distant past and future, the present moment and place are present at the same time. It is impossible to think of only one concept without not thinking of the other. Derrida's \cite{b58} différance describes a similar phenomenon. The meaning of a word can change nuance of meaning several times in the context of reading a piece of writing. Different associations, possible understanding, inevitable misunderstanding, and last but not least ambiguity arise. Eisenman \cite{b59} describes this situation in the context of absence, memory, and immanence. In this linguistic model, memory and immanence, absence and presence are always mutually dependent.

"Absence is either the trace of a former presence, in which case it involves memory, or it is the trace of a possible presence, in which case it has immanence." \cite{b59}

Ambiguous illusion also arises from different interpretations of the same object. In this context, Stuart Franklin \cite{b50} cites the superimposition of contradictory visual codes or deliberate genre blends that reinforce the ambiguity of the image's message. As an example, he cites René Magritte's disorienting painting La Condition Humaine \cite{b60}, in which the canvas morphs into the landscape, and a rendering of Edward Weston's Cabbage Leaf  \cite{b61}, in which the image oscillates between two or more interpretations: is it a folded fabric or a simple cabbage leaf?

The pictorial illusion is also fostered by what is called mimetic desire, by the viewer's prior knowledge and desires. This phenomenon is based on the emotional relationship between the viewer and the subject of the image. This also includes the play with culture-specific imprints and narratives. The play with so-called memes or Easter Eggs also belongs to this category. These are iconic images that we all know from the media and that have burned themselves into our collective memory. Ambiguity also arises from the use of culture-dependent psychological or emotional messages, as we know them from product advertising. The illustrators of Jules Verne \cite{b22, b23, b24} are fond of using mimetic desire in the overlaying of codes and in deliberate decontextualization. Very specifically, an illustration in the book Journey Around the Moon is very reminiscent of Raffael's "Transfiguration" \cite{b18}. The illustration shows the three spacemen in a weightless configuration, similar to Raffael's. The allusion to the famous model is clear. Dreams of Dalí in Virtual Reality \cite{b13} also play with the superimposition of ambiguous codes. We recognize spider legs and elephant-like animals, combined into hybrid hybrid beings. A desert landscape gives birth to ambiguous figures.

Discontinuity and decontextualization, deliberate omission of facts and the superimposition of contradictory codes, topology and a certain indeterminacy-these are tangible methodological suggestions for game designers to create ambiguous utopian worlds.

\newpage
\section {Computer Generated Illustrations}

"I prefer people to look at my pictures and invent their own stories." - Josef Koudelka \cite{b50}.\\

The section that follows now shows six computer-generated image compositions. For my works I like to use and have been using the open source software Blender for almost twenty years. In the following works six different utopian scenes are presented. They are created as virtual worlds to selected quotes from literature. They are three-dimensional, de- and re-contextualized illustrations. These images address ambiguity in utopian representation through discontinuity and decontextualization, the omission of facts, and the superimposition of contradictory codes. The color computer graphics in this chapter show sketches of six fantastic spaces and sceneries, with no specific context, no indication of location and often without any scale. They are re-contextualized with selected quotes from literature. These are quotes from Herbert George Wells, Stefan Zweig, Max Frisch, Thomas Morus and Friedrich von Schiller. Through this re-combination and re-contextualization emergent individual interpretations of real non-existent spaces are created. Places without time, between dream and reality. Places that may never have existed without this current pictorial manifestation and will never exist in the future. They are integrated into a new context and remain open spaces for the dreams, fantasies, desires and associations of the viewer in the brief, present moment of encounter. In this brief present, a utopia is created in our own consciousness. What associations arise in you when you look at these pictures?

\subsection{A Cool Vista of Blue and Purple}

\begin{figure*}[htbp]
\begin{minipage}[b]{1.0\textwidth}
\centerline{\includegraphics[width=1.0\textwidth]{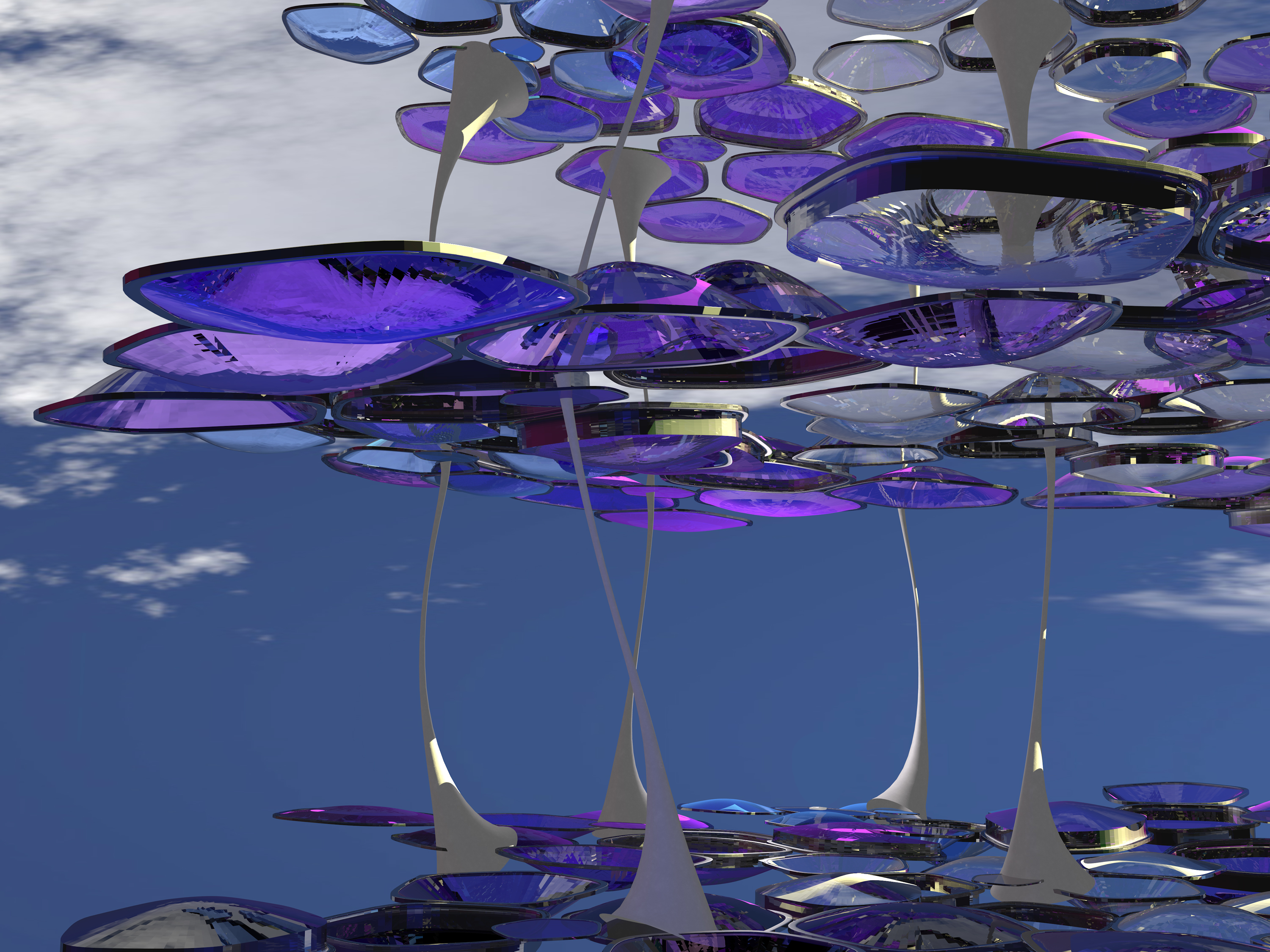}}
\caption{A Cool Vista of Blue and Purple © 2014 Wolfgang Höhl, München.}
\label{fig4}
\end{minipage}
\end{figure*}

Figure \ref{fig4} shows a first illustration to a quote from "When The Sleeper Wakes" by H. G. Wells \cite{b62}:\\

"The passage ran down a cool vista of blue and purple, and ended remotely in a railed space like a balcony, brightly lit and projecting into a space of haze, a space like the interior of some gigantic building. Beyond and remote were vast and vague architectural forms."

\subsection{The Awakening}

\begin{figure*}[htbp]
\begin{minipage}[b]{1.0\textwidth}
\centerline{\includegraphics[width=1.0\textwidth]{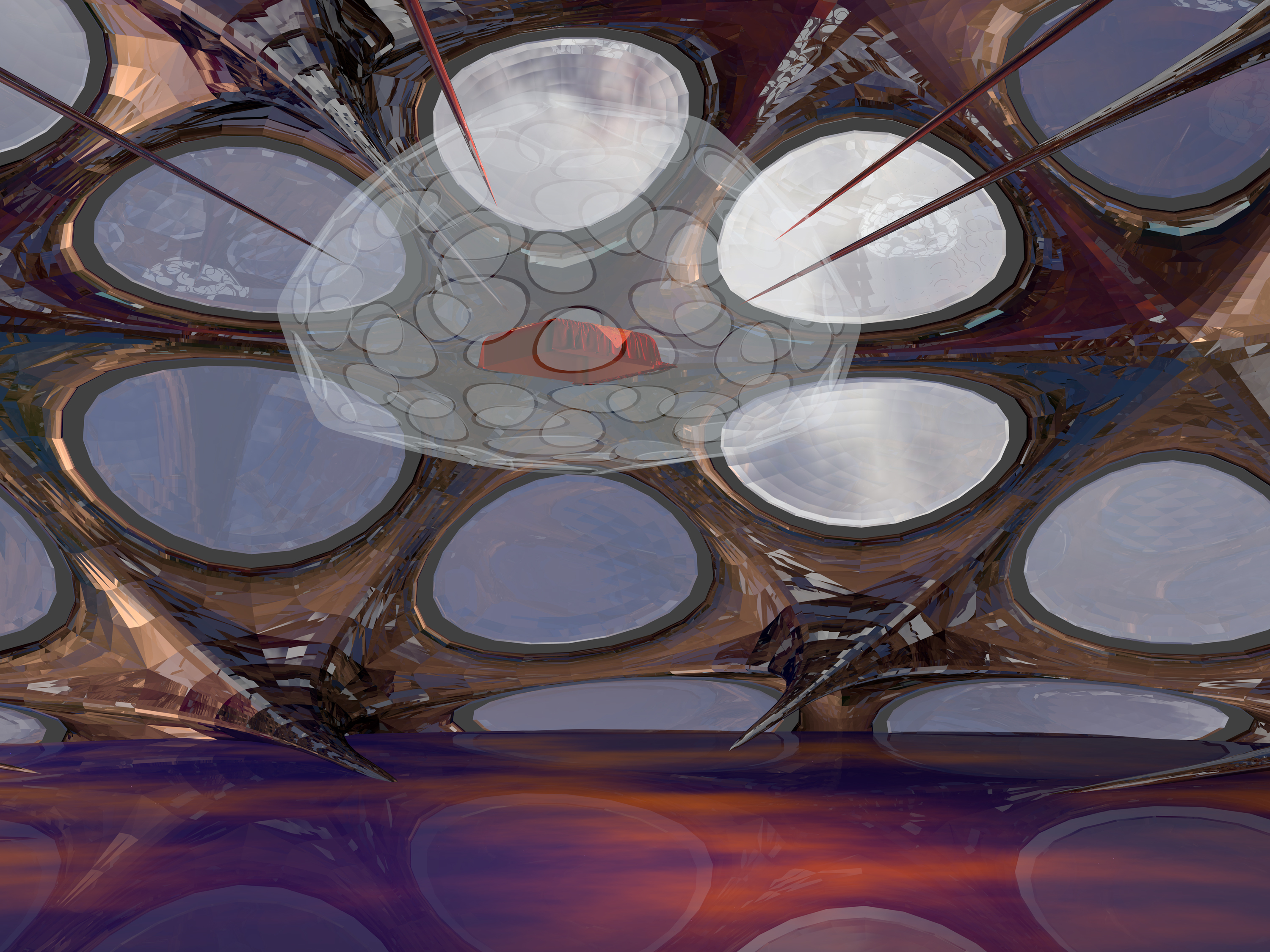}}
\caption{The Awakening © 2014 Wolfgang Höhl, München.}
\label{fig5}
\end{minipage}
\end{figure*}

Figure \ref{fig5} shows a second illustration to another quote from "When The Sleeper Wakes" by H. G. Wells \cite{b62}:\\

"The slightly greenish tint of the glass-like substance which surrounded him on every hand obscured what lay behind, but he perceived it was a vast apartment of splendid appearance, ... "

\subsection{Brennendes Geheimnis}

\begin{figure*}[htbp]
\begin{minipage}[b]{1.0\textwidth}
\centerline{\includegraphics[width=1.0\textwidth]{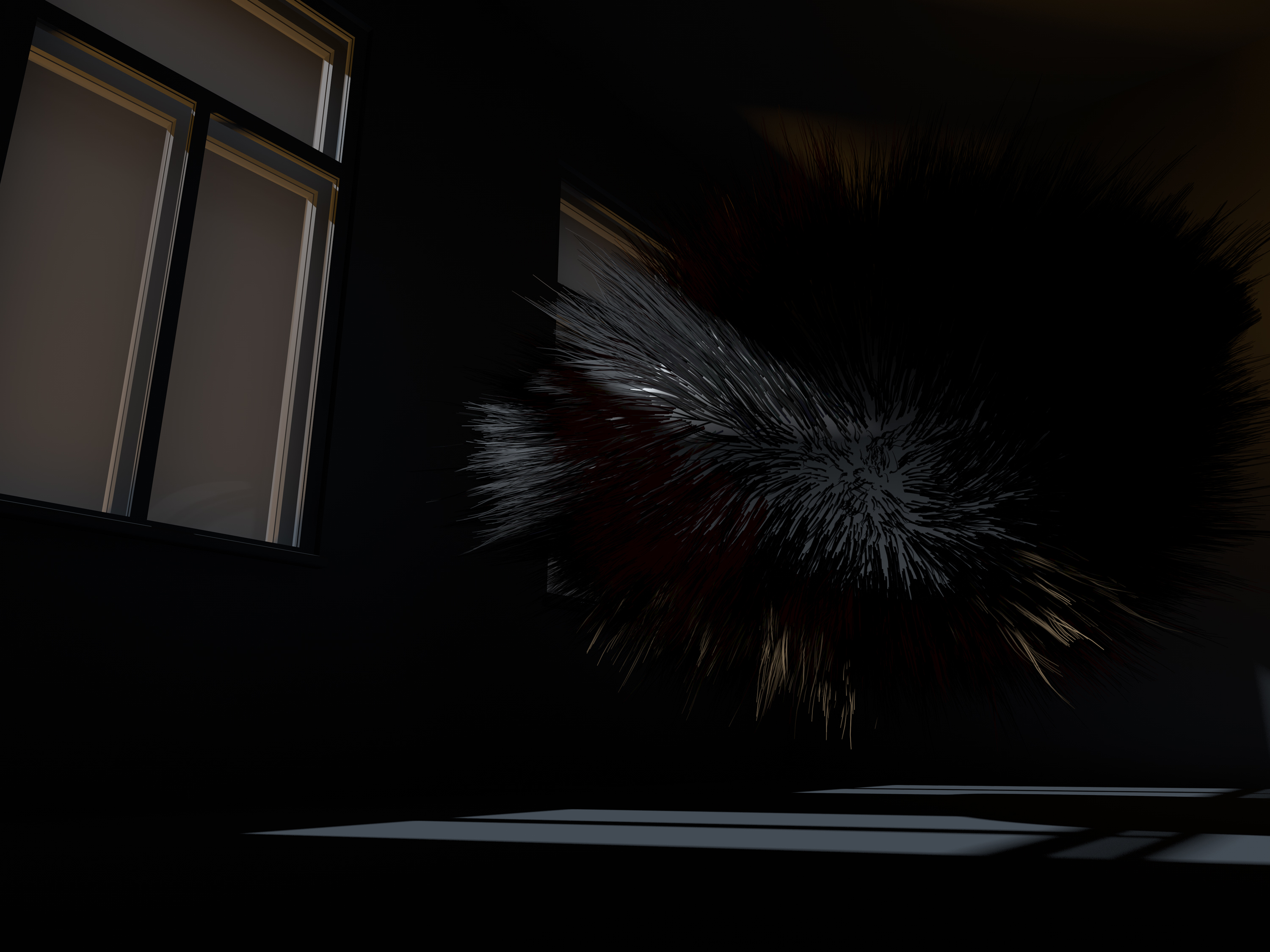}}
\caption{Brennendes Geheimnis © 2014 Wolfgang Höhl, München.}
\label{fig6}
\end{minipage}
\end{figure*}

The following lines are taken from "Brennendes Geheimnis" (Burning Secret) by Stefan Zweig \cite{b63}. Figure \ref{fig6} shows the referring illustration.\\

"Das alles war sehr süß und schmeichlerisch nun im Dunkel zu denken, leise schon verworren mit Bildern aus Träumen, und beinahe war es schon Schlaf. Da war ihm, als ob plötzlich die Türe ginge und leise etwas käme. Er glaubte sich nicht recht, war auch schon zu schlafbefangen, um die Augen aufzutun."\\

"All this was very sweet and flattering now to think in the dark, quietly already tangled with images from dreams, and almost it was already sleep. He felt as if suddenly the door opened and something came quietly. He did not quite believe himself, was already too caught up in sleep to open his eyes."

\subsection{Homo Faber}

\begin{figure*}[htbp]
\begin{minipage}[b]{1.0\textwidth}
\centerline{\includegraphics[width=1.0\textwidth]{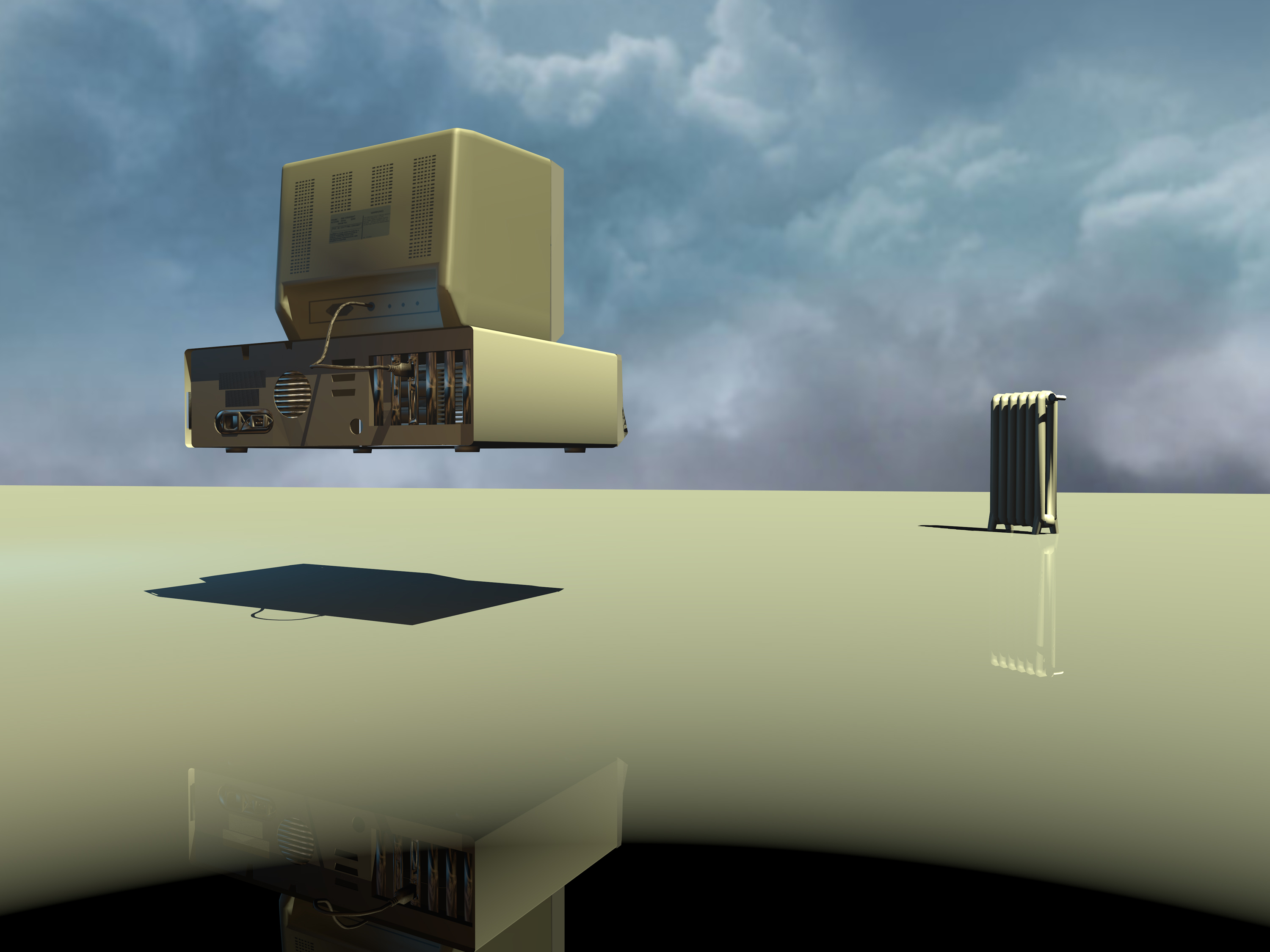}}
\caption{Homo Faber © 2014 Wolfgang Höhl, München.}
\label{fig7}
\end{minipage}
\end{figure*}

Figure \ref{fig7} shows an illustration to the following quote from "Homo Faber" by Max Frisch \cite{b64}:\\

"Verdreifachung der Menschheit in einem Jahrhundert. Früher keine Hygiene. Zeugen und gebären und im ersten Jahr sterben lassen, wie es der Natur gefällt, das ist primitiver, aber nicht ethischer."\\

"Tripling of humanity in one century. In the past, no hygiene. To witness and give birth and let die in the first year as nature pleases, that is more primitive, but not more ethical."

\subsection{Das verschleierte Bild zu Sais}

\begin{figure*}[htbp]
\begin{minipage}[b]{1.0\textwidth}
\centerline{\includegraphics[width=1.0\textwidth]{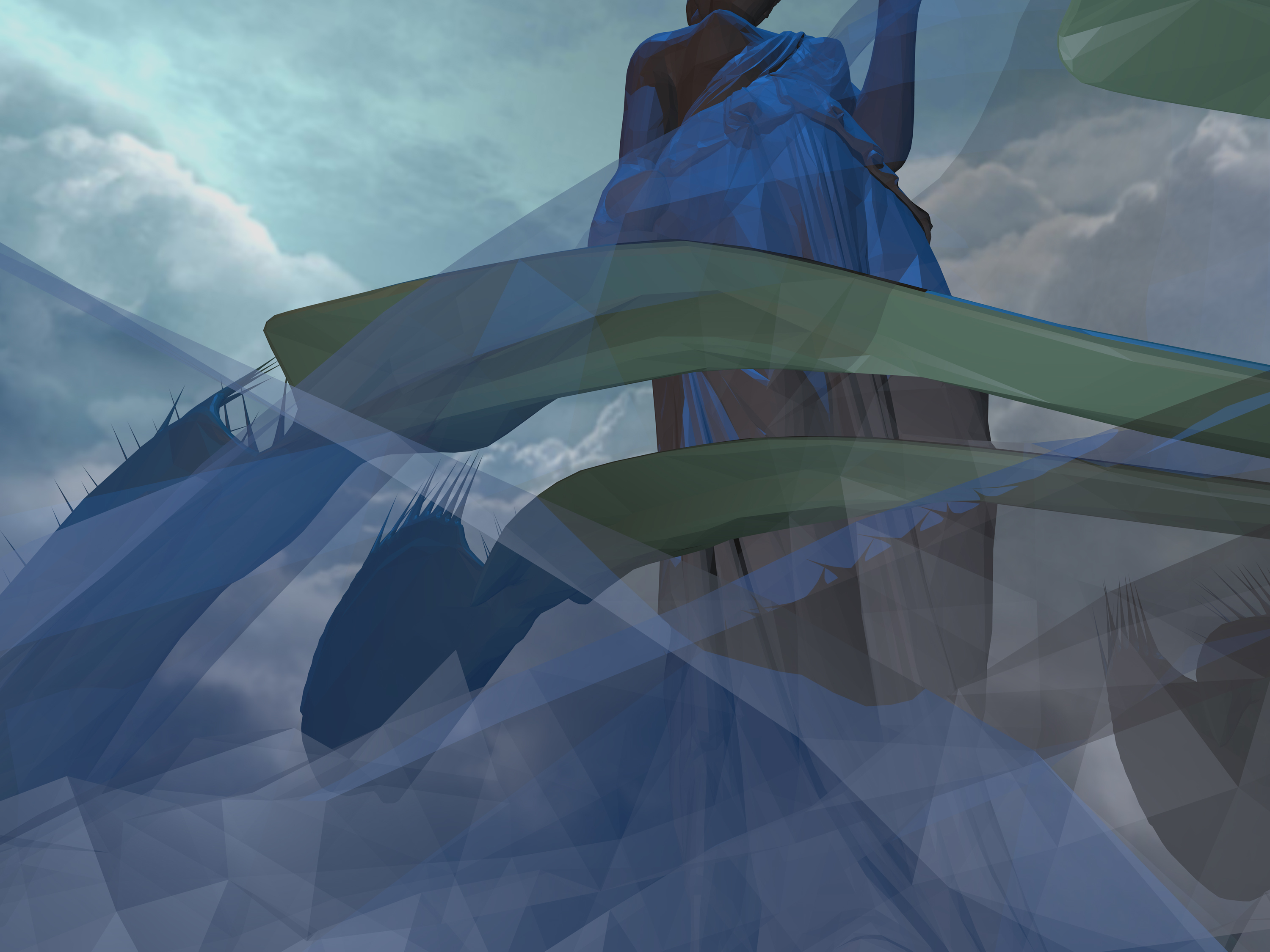}}
\caption{Das verschleierte Bild zu Sais © 2014 Wolfgang Höhl, München.}
\label{fig8}
\end{minipage}
\end{figure*}

Friedrich Schillers lines from "Das verschleierte Bild zu Sais" (The veiled image to Sais) \cite{b65} are placed in the context of Figure \ref{fig8}.\\

“Von oben durch der Kuppel Oeffnung wirft\newline
Der Mond den bleichen silberblauen Schein\newline
Und furchtbar wie ein gegenwärt´ger Gott\newline
Erglänzt durch des Gewölbes Finsternisse\newline
In ihrem langen Schleier die Gestalt.”\\

"From above through the dome opening\newline
The moon casts the pale silver-blue glow\newline
And terrible as a present god\newline
Shining through the darkness of the vault\newline
Appears the figure in its long veil."

\subsection{Utopia}

\begin{figure*}[htbp]
\begin{minipage}[b]{1.0\textwidth}
\centerline{\includegraphics[width=1.0\textwidth]{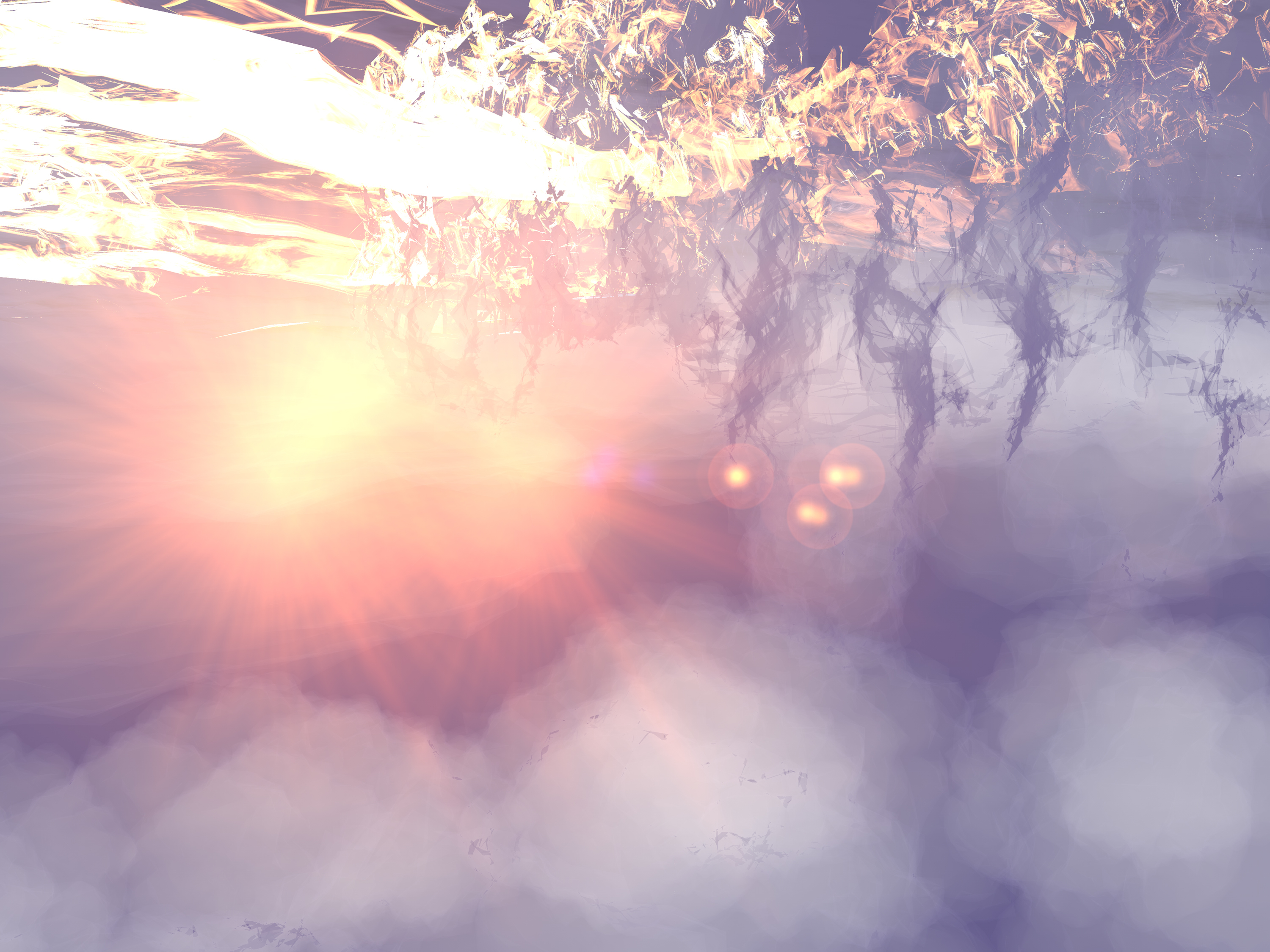}}
\caption{Utopia © 2014 Wolfgang Höhl, München.}
\label{fig9}
\end{minipage}
\end{figure*}

In "Utopia" Thomas Morus describes a curious society living on a strange island \cite{b19, b20}. One quote from his work will be juxtaposed to Figure \ref{fig8}.\\

"Situm est igitur Amaurotum in leni deiectu montis. figura fere quadrata. Nam latitudo eius paulo infra collis incoepta uerticem millibus passuu duobus ad flumen Anydru pertinet, secundum ripam aliquanto longior." 

\newpage
\section{Utopia as a Realm for New Ideas}

It has been shown that the three design criteria for ambiguous pictorial representation are practical and applicable. Discontinuity and decontextualization, the deliberate omission of facts, and the superimposition of contradictory codes are helpful principles in the design and implementation of utopian representations. In practical work on six computer-generated illustrations, the above design criteria could be successfully applied to the design of virtual worlds.

An ontology of interactive media can represent important correlations between users, content and technology. It is also able to explain the elusive "visual poetry" and image perception. Figure \ref{fig2} and Figure  \ref{fig3} can be used to accurately define and assign terms for game design. Popper's 3-world model \cite{b1} can be seamlessly integrated and is a good complement to the presented five-part model of media ontology. The intrinsic embeddedness of the user's physical, psychological, and spiritual reality in the media system is clearly demonstrated.

The overview of utopian representation shows that utopia is a cross-media phenomenon. It has also clarified that utopias are always created against the background of current social events. They always refer to an existing, concrete reality. Thus, game design is also always in an interdisciplinary and social context. The sustainable design of virtual worlds can make a significant contribution to developing powerful visions and better shaping our future reality.

We can develop many positive utopias today. Climate change, energy transition and growing social inequality urgently need positive utopias and the meaningful, intelligent and sustainable technical innovation. We also need to examine our current economic and social order for future viability. "Social Market Economy in the Digital Future" is the title of the current Foresight short report commissioned by the German Federal Ministry of Economics \cite{b66}. 

This study shows us the way to think about sustainable, alternative and complementary concepts to an outdated, growth-oriented economy, to formulate and implement reforms. There is little time. And we have known this since the early 1970s  \cite{b67}. What cities and regions do we want to live in? Smart cities and a comprehensive sustainable digital transformation need guiding principles and higher digital literacy. New automated applications with artificial intelligence are worthless without a well-founded and socially broad-based ethical discussion. There is plenty of room for visionary ideas and a better future. In an interview with Ezra Klein in the New York Times, former U.S. President Barack Obama said:\\

" ... the differences we have on this planet are real. They're profound. And they cause enormous tragedy as well as joy. ... We're just a bunch of humans with doubts and confusion. We do the best we can. And the best thing we can do is treat each other better because we're all we've got. ... " \cite{b68}.

\newpage
\section*{Acknowledgment}

This paper was developed as a publication on my presentation at the Science MashUp Conference on the topic of "XR - Society - Utopia" at Leipzig University of Applied Sciences on April 24, 2021. My sincere gratitude goes to Gabriele Hooffacker at Leipzig University of Applied Sciences, who initiated and organized this event. Last but not least, my thanks go to my partner Cornelia for her sustained support.

\newpage

\begin{figure}[htbp]
\includegraphics[width=0.35\columnwidth]{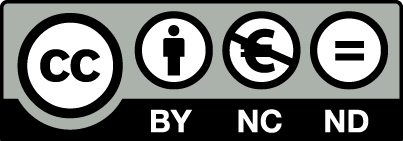}
\end{figure}

© 2021 by the author. This publication is licensed under the terms and conditions of the Creative Commons Attribution (CC BY-NC-ND) license (https://creativecommons.org/licenses/by-nc-nd/4.0/).


\end{document}